\documentclass[a4paper,fleqn]{cas-dc}

\usepackage[authoryear]{natbib}
\def\tsc#1{\csdef{#1}{\textsc{\lowercase{#1}}\xspace}}
\tsc{JUN}
\begin{document}
\let\WriteBookmarks\relax
\def\floatpagepagefraction{1}
\def\textpagefraction{.001}
\shorttitle{Complex SSS spectra}
\shortauthors{J.-U. Ness}

\title [mode = title]{The complications of learning from
Super Soft Source X-ray spectra}
%\tnotemark[1,2]

%\tnotetext[1]{This document is the results of the research
%   project funded by the National Science Foundation.}

%\tnotetext[2]{The second title footnote which is a longer text matter
%   to fill through the whole text width and overflow into
%   another line in the footnotes area of the first page.}

\author[1]{Jan-Uwe Ness}[type=editor,
                        auid=000,bioid=1,
                        prefix=Dr.,
                        role=Researcher,
                        orcid=0000-0003-0440-7193]
%\cormark[1]
\ead{juness@sciops.esa.int}
\ead[url]{januweness.eu, juness@sciops.esa.int}

\address[1]{XMM-Newton and Integral SOC, European Space Astronomy Centre, Camino Bajo del Castillo s/n, Urb. Villafranca del Castillo, 28692 Villanueva de la Canada, Madrid, Spain}

\cortext[cor1]{Corresponding author}

\begin{abstract}
Super Soft X-ray Sources (SSS) are powered by nuclear burning on the surface
of an accreting white dwarf, they are seen around 0.1-1\,keV (thus in
the soft X-ray regime), depending
on effective temperature and the amount of intervening interstellar
neutral hydrogen ($N_{\rm H}$). The most realistic model
to derive physical parameters from observed SSS spectra would be an
atmosphere model that
simulates the radiation transport processes. However, observed
SSS high-resolution grating spectra reveal highly complex details
that cast doubts on the feasibility of achieving unique results from
atmosphere modeling. In this article, I discuss two independent
atmosphere model analyses of the same data set, leading to different
results. I then show some of the details that complicate the analysis
and conclude that we need to approach the interpretation
of high-resolution SSS spectra differently. We need to focus more on
the data than the models and to use more phenomenological approaches
as is traditionally done with optical spectra.
\end{abstract}

%\item Research highlights item 1
%\item Research highlights item 2
%\item Research highlights item 3


\maketitle

\section{Introduction}

Cataclysmic Variables are characterised by mass transfer from a
companion star onto the surface of a white dwarf. If there was only
accretion, the white dwarf would grow in mass until reaching
the Chandrasekhar mass limit, when it either collapses into a neutron
star or explodes as a supernova Ia (if the composition is rich in
carbon and oxygen). However, life is more complicated. Since the
accreted material is rich in hydrogen, it can fuse to helium, an
energy source that leads to mass loss which can be more than the
amount of mass gained by accretion. This may happen after a long episode
of accretion leading to a successive increase in temperature and
pressure. If ignition conditions are reached, the accreted hydrogen
explodes in a thermonuclear runaway known as a nova. The radiation
pressure drives an initially optically thick wind that obscures
any high-energy radiation from the nuclear burning zones. If the
mass lost in the wind is higher than the mass previously accreted,
then the white dwarf cannot reach the Chandrasekhar mass limit. When
the wind becomes optically thin, the central burning regions can
be seen. Radiation temperatures then reach several $10^5$\,K, and
a blackbody of this temperature has it's Wien tail in the soft
X-ray regime, around 1\,keV ($\sim 10$\,\AA). A small fraction
of white dwarfs permanently host conditions of temperature
and pressure to allow the hydrogen-rich accreted material to undergo
nuclear fusion at the same rate as the accretion rate.
The observational class of these Super Soft Sources (SSS)
was discovered by the {\it Einstein} satellite and
significantly expanded with large-area {\it ROSAT} observations.
Initially it was believed that they were powered by accretion in neutron
stars, but \cite{heuvel} proposed nuclear burning on white dwarfs
which was later confirmed by \cite{kahab}.\\

In order to see a permanent SSS spectrum, special fine tuning of
several parameters such as accretion rate and burning rate is
needed, and the class is thus very small. Further, the softness
of the spectra allows us to only see them along lines of sight
with low $N_{\rm H}$, limiting the sample to either nearby systems
or extragalactic ones such as Cal\,83 in the LMC.\\

Transient SSS emission can be produced during nova outbursts which
have been observed with much higher count rates. It is conceivable
that novae may also be intrinsically more luminous than permanent SSS
because they rely on a much larger reservoir of previously accreted
and accumulated material. During the early phases of a nova outburst,
the high-energy
radiation produced on the surface of the white dwarf is blocked
by higher layers of optically thick material that has been ejected
by radiation pressure. Generally, after several weeks to months, the
ejecta become optically thin, and the soft X-ray emission can be
observed. If, however, nuclear burning switches off before the
ejecta have cleared to allow SSS emission to escape, the nova
may never be seen as a transient SSS. Some examples are discussed by
\cite{schwarz2011}.\\

Transient SSS emission can be observed in other galaxies, owing to
their brightness and the low amount of interstellar absorption.
As example is the remarkable short-period recurrent nova in
M31N2008-12a \citep{henze14} with multiple outbursts having bee
observed every year since 2013. Observations of SSS emission from
other galaxies may be easier than for Galactic novae because of
the lower amount of interstellar absorption.\\

An early example of transient SSS emission from a nova outburst was
described by \cite{krautt96} based on {\it ROSAT} data (0.1-2.4\,keV).
The spectral resolution of $R=\Delta$E/E$=0.5$ at 1\,keV is low enough that a
blackbody fit already reproduced the data. However, the resulting
bolometric luminosity, derived from the Stefan-Boltzmann law
assuming spherical symmetry, was unrealistically high (far above
the Eddington limit), and \cite{krautt96} emphasised that blackbody
fits do not lead to reliable parameter estimates. A followup paper
by \cite{balm98} analysed the same data set with LTE atmosphere
models, also yielding good fits to the data but lower luminosities.
The more realistic physical assumptions led the authors to conclude
that more realistic results have been found. A similar analysis
was performed by \cite{parmarcal83,parmarcal87} using BeppoSAX data
(0.1-10\,keV, $R=0.18$ at 1\,keV using equation 10 in \citealt{parmar97})
of the two
prototype SSSs Cal\,83 and Cal\,87, demonstrating that atmosphere
models lead to lower luminosities, however, not to better values
of $\chi^2$. The spectral resolution of BeppoSAX was designed to
resolve absorption edges which are the most striking features
at the temperatures and energies of SSS spectra. Abundances estimates
are thus possible with the non-dispersive spectrometers.\

\section{High-resolution X-ray spectra of Super Soft Sources}

\begin{figure}
%        \centering
        \includegraphics[scale=.35]{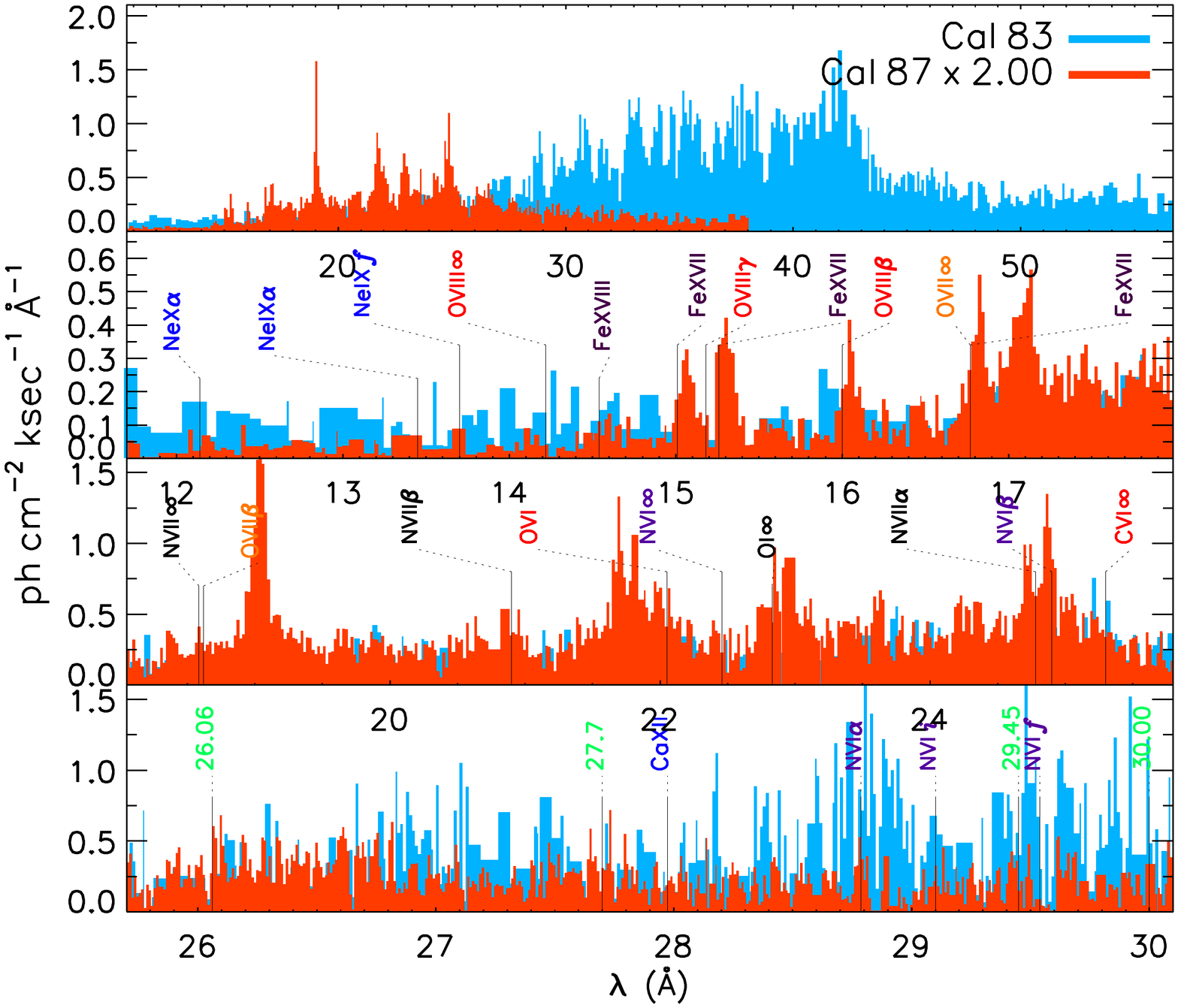}
        \includegraphics[scale=.30]{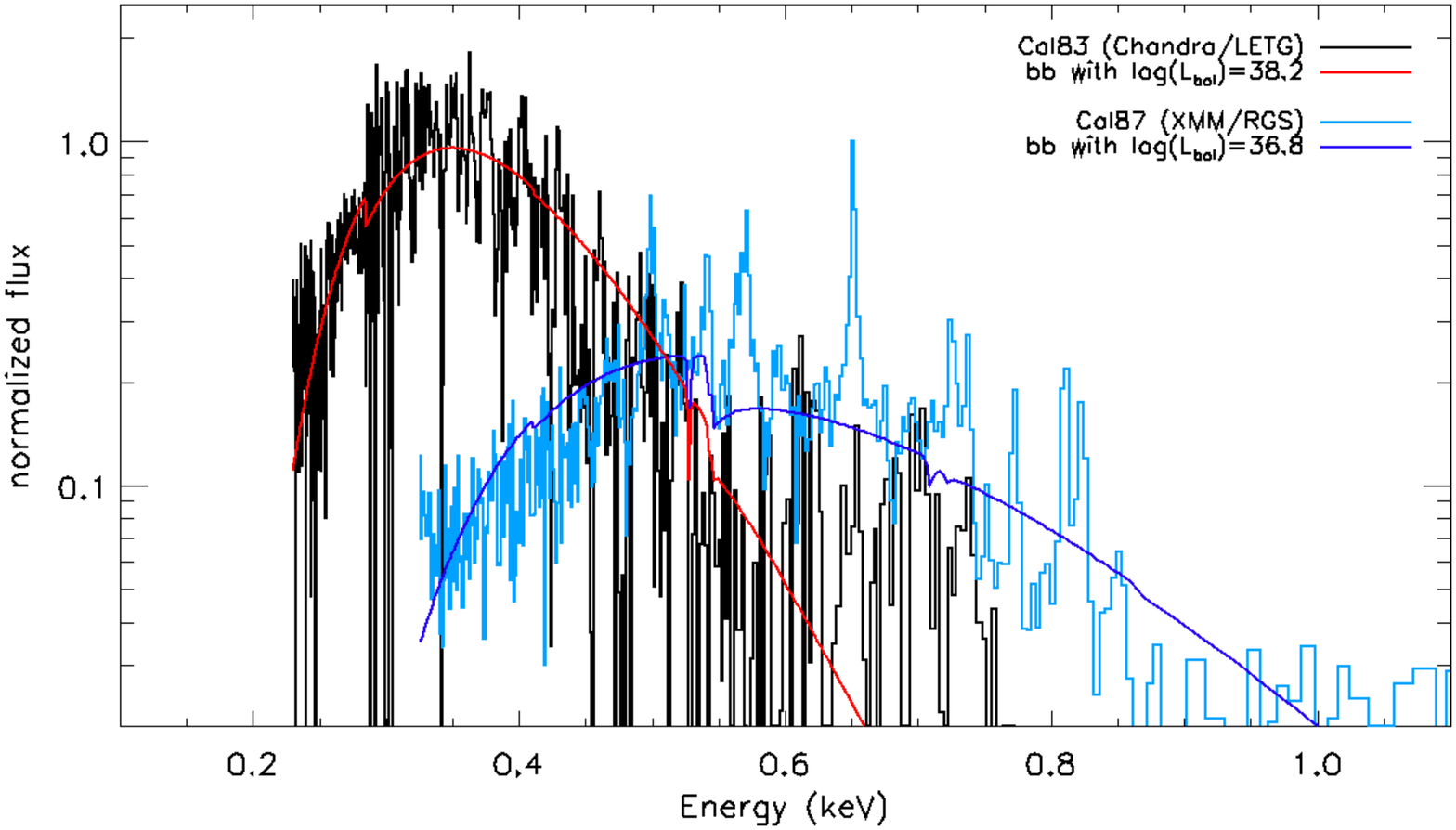}
        \caption{Comparison of high-resolution X-ray
	spectra of the prototype SSSs Cal\,83 ({\it Chandra}
	ObsID 1900)
	and Cal\,87 ({\it XMM-Newton} ObsID 0153250101), both
        located in the LMC. In the bottom panel, blackbody fits
        are used to illustrate that Cal\,83 is an atmosphere
        spectrum while Cal\,87 hosts strong emission lines on
        top of blackbody-like continuum.}
        \label{c8387}
\end{figure}

With the advent of the high-resolution X-ray gratings, operated
on board the {\it XMM-Newton} ($R=0.004$ at 1\,keV) and
{\it Chandra} ($R=0.001$ at 1\,keV) missions, it
became possible to resolve important details that are needed to
better constrain atmosphere models, such as elemental
abundances from the depth of absorption lines. From low-resolution
spectra, only very rough abundance patterns can be estimated from
the depth of absorption {\em edges}. The first X-ray grating spectra
of SSSs were taken of the prototype SSSs Cal\,83 and Cal\,87,
and a comparison of these spectra is shown in Fig.~\ref{c8387}.
Even though, both sources are at the same distance, they differ
in flux by a factor two (although Cal\,87 is slightly brighter
at higher energies), but more strikingly, the
spectrum of Cal\,87 is an emission line spectrum while Cal\,83
resembles more an atmosphere spectrum which was successfully
modeled with a non-LTE atmosphere model by \cite{lanz04}.
The low-resolution spectrum of Cal\,87, shown by \cite{parmarcal87},
is more structured, but the data quality is low enough to consider
the structure as Poisson noise. A blackbody fit and an
atmosphere model fit still yield good fits to the poor data while
the high-resolution spectra now show that both models are
wrong.
A very similar grating spectrum to that of Cal\,87 was found
by \citep{nessusco} with the recurrent nova U\,Sco that is an
eclipsing system, leading \cite{nessobsc}
to conclude that SSS spectra like that of Cal\,87 (which is also
eclipsing) are strongly
affected by obscuration effects, e.g., by an accretion disc in
a high-inclination system. In the bottom panel of Fig.~\ref{c8387},
blackbody fits illustrate that for Cal\,87, there is weak
underlying atmospheric continuum emission, possibly originating
from the surface, a fraction of which getting to the observer via
Thomson Scattering \citep{nessusco}. Assuming the bolometric
luminosity within reasonable limits, the observed effective
temperature (Wien tail) requires a radius of the pseudo photosphere
to be small enough that it must be fully eclipsed by the companion
around phase 0. The fact that we see continuum emission requires the
continuum emission to have undergone some scattering processes.
The fact that the continuum is not spectrally distorted indicates
that the scattering mechanism is independent of photon energy, thus
the conclusion by \cite{nessusco} that we are dealing with Thomson
scattering.\\

\section{Analysing high-resolution SSS spectra with atmosphere models}

\begin{figure}
%        \centering
        \includegraphics[scale=.30]{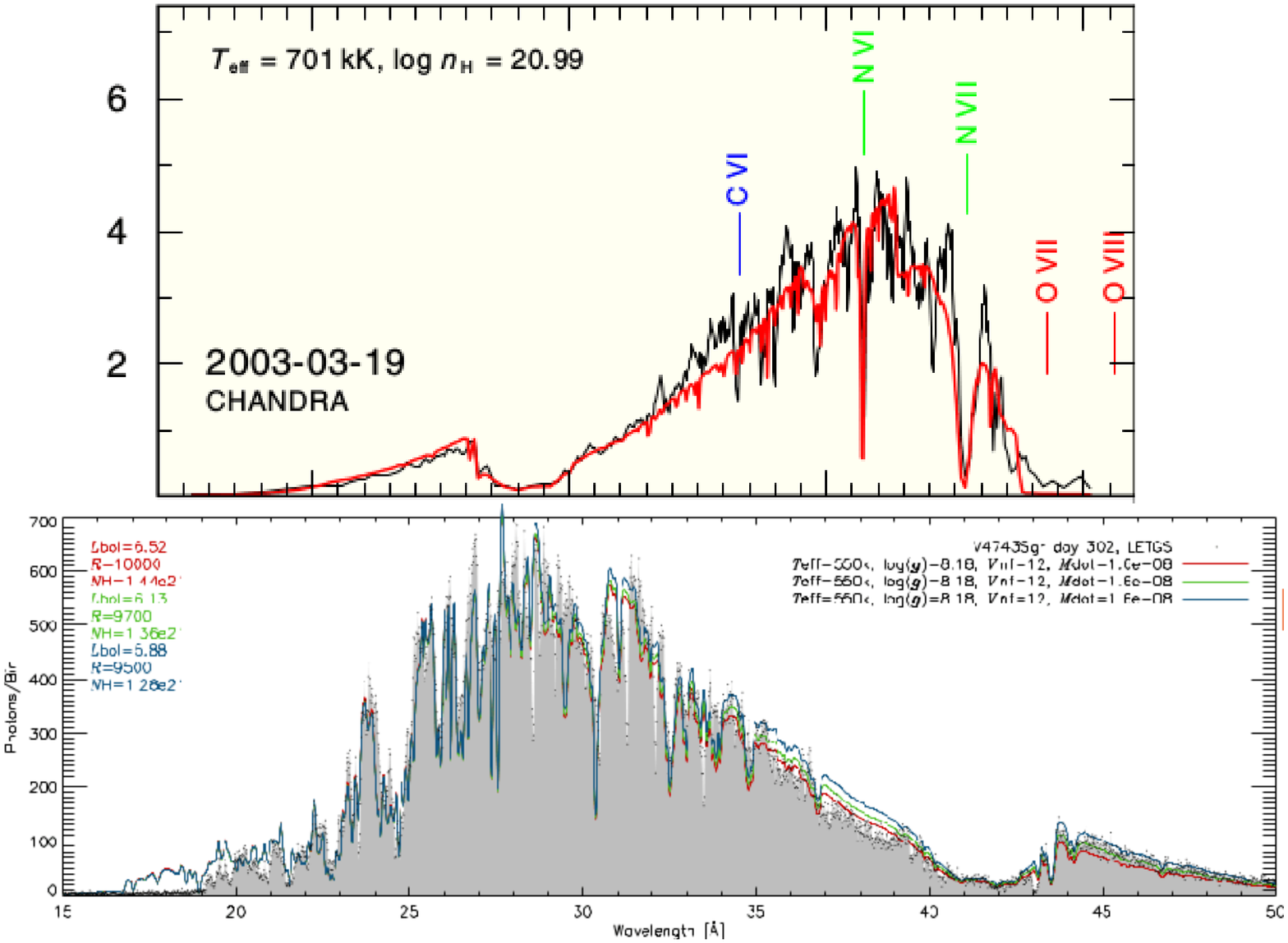}
        \caption{Two independent approaches for fitting atmosphere
	models to the same data set (Nova V4743\,Sgr, day 180 after
	outburst, {\it Chandra} ObsID 3775 (day 180)) by
	\cite{rauch10} (top) and
        by \cite{vanrossum2012} (bottom).}
        \label{cmpatm}
\end{figure}

After \cite{krautt96} made a strong case that blackbody fits are
not reliable, several approaches were attempted to use atmosphere
models. In the most central core of an atmosphere model, there is
still a blackbody model (source function), and the Stefan-Boltzmann
law also applies to atmosphere models - at least on a qualitative
level. While the blackbody model makes the extremely simplifying
assumption of thermal equilibrium (TE), the next step foreward is to
assume local thermal equilibrium (LTE, e.g., \cite{balm98}).
More modern codes compute non-LTE (NLTE) atmospheres. One such code
developed for white dwarfs is the plane-parallel hydrostatic
atmosphere code TMAP (T\"ubingen NLTE Model-Atmosphere Package).
This code has proven particularly useful for compact objects such as
isolated white dwarfs, e.g. \cite{werner12} or neutron stars, e.g.
\cite{rauch08}. In the limit of highly compact objects, a plane
parallel geometry yields similar results as a spherically symmetric
geometry which is computationally more expensive. A second important
atmosphere code is the PHOENIX code by P. Hauschildt that solves
the NLTE radiative transport equations in a co-moving frame, thus
allowing expanding ejecta to be modelled. Based on PHOENIX, a
wind-type model for SSS spectra has been developed by \cite{vanrossum2012}.\\

\begin{figure}
%        \centering
        \includegraphics[scale=.28]{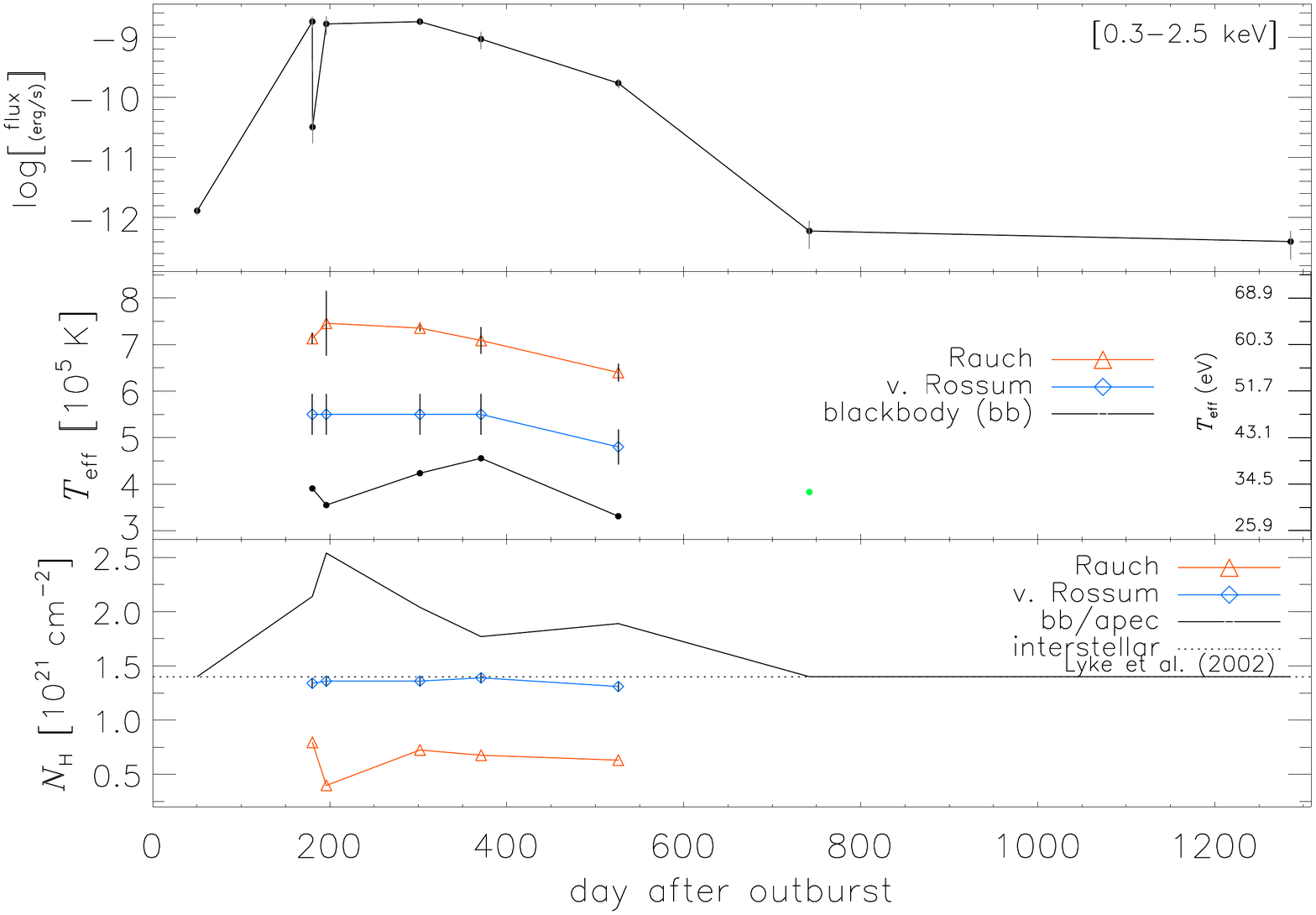}
        \caption{Comparison of atmosphere model parameters of the
        two models by \cite{rauch10} and by \cite{vanrossum2012}.
	The observed (absorbed) fluxes shown in the top have been
	derived by integration
        of the fluxes in each wavelength bin.}
        \label{atmdiff}
\end{figure}

The first nova observed as a transient SSS with an X-ray grating
was V4743\,Sgr \citep{v4743}, and five SSS spectra where taken at
different times during the evolution of this outburst. This dataset allows detailed
conclusions about the evolution of the ejecta during the SSS phase.
Two independent analyses with atmosphere models were performed by
\cite{rauch10} (based on TMAP) and
by \cite{vanrossum2012} with examples of their models compared to
data shown in Fig.~\ref{cmpatm}. One can see that agreement
with the overall shape has been achieved in both cases while there
are substantial differences in the details. No statistical goodness
criterion like $\chi^2$ was given, and from a statistical point of
view, both fits are unacceptable. In Fig.~\ref{atmdiff}, I compare
the evolution of the parameters effective temperature (middle panel)
and $N_{\rm H}$ (bottom) derived from the two approaches and a
simple blackbody fit. The effective temperature systematically differs
by $\sim 25$\%. \cite{vanrossum2012} has assumed a constant value
of $5.5\times 10^5$\,K while \cite{rauch10} have iterated this
parameter, detecting small variations above $7\times 10^5$\,K.
A blackbody fit yields the lowest effective temperature values.
The trends in temperature evolution differ only slightly. The
TMAP model shows a small increase in temperature until day 300
and a continuous decline thereafter. Meanwhile, the wind-type model
can explain the changes that the TMAP model attributes to temperature
changes to changes in other parameters and thus concludes the
observations to be consistent with constant effective temperature
until day 370 and only then a decrease. The blackbody fits yield
the largest changes in best-fit effective temperature which is
likely a result of ignoring any absorption that may be variable.\\

Shortly after the outburst, the amount of interstellar absorption
by neutral hydrogen was determined by \cite{lyke} with
$1.4\times 10^{21}$\,cm$^{-2}$, based on a method by
\cite{ghj74} that was also employed by \cite{Gehrz2015} for
the nova V339\,Del, also giving some more details. Interestingly,
exactly the same value was found by \cite{vanrossum2012} who
argues that $N_{\rm H}$ requires special
attention prior to any adjustment of atmosphere model parameters.
Small variations in $N_{\rm H}$ imply large differences in assumed
fluxes at long (EUV) wavelengths at which hardly any flux
is observed. Therefore, poor values of $N_{\rm H}$ will hardly be
penalised by $\chi^2$ fitting even though strong assumptions
are made about the invisible part of the spectrum. After
careful pre-adjustment of $N_{\rm H}$, the values determined
by \cite{vanrossum2012} are all spot on with those derived by
\cite{lyke}. I emphasise here that \cite{lyke} was not quoted
by \cite{vanrossum2012}, indicating that \cite{vanrossum2012}
was not biased by the
literature value. Meanwhile, the $N_{\rm H}$ values derived by
\cite{rauch10}, determined by fitting $N_{\rm H}$ simultaneously
with other
atmosphere parameters, are all well below the value derived
by \cite{lyke} (which was also not quoted).
This can either mean that \cite{lyke} overestimated $N_{\rm H}$,
that the value of $N_{\rm H}$ was lower during the SSS phase, or
that \cite{rauch10} underestimate $N_{\rm H}$. Meanwhile, the values
of $N_{\rm H}$ from the blackbody fits are all much higher which can
either mean that local absorption was higher during the SSS phase or
that the blackbody fit overestimates $N_{\rm H}$.
As all of the spectral fits have unacceptable goodness of fit,
it is also possible that all of the derived parameter values are
incorrect.\\

The reason why the effective temperature can be underestimated
while $N_{\rm H}$ is overestimated can be explained as follows:
\begin{itemize}
	\item If the ionisation energy of an abundant element such
		as nitrogen coincides with the Wien tail, a
		significant amount of high-energy emission is missing.
	\item A model not accounting for ionisation absorption edges
		will see the Wien tail at longer wavelengths, thus
		underestimating the effective temperature.
	\item A model with underestimated effective temperature will
		overpredict the emission at soft energies, and
		rigorous parameter fitting will increase $N_{\rm H}$
		in order to get rid of the excess soft emission.
\end{itemize}
Inversely, if the effects of absorption edges in the Wien tail are
overestimated, then effective temperatures are overestimated and
$N_{\rm H}$ underestimated. The large differences between the
two atmosphere models in $N_{\rm H}$ can thus be a consequence of the
difference in effective temperature, and it is of high importance that
the absorbing behaviour of the outer ejecta is well understood in
order to derive reliable principle parameters.

\section{The annoying little details}

In the following subsections, a few effects are described that are
seen in the observed spectra that complicate any quantitative
analysis based on global models.

\subsection{Blue shifts}
\label{S:BlueShift}

\begin{figure}
%        \centering
        \includegraphics[scale=.28]{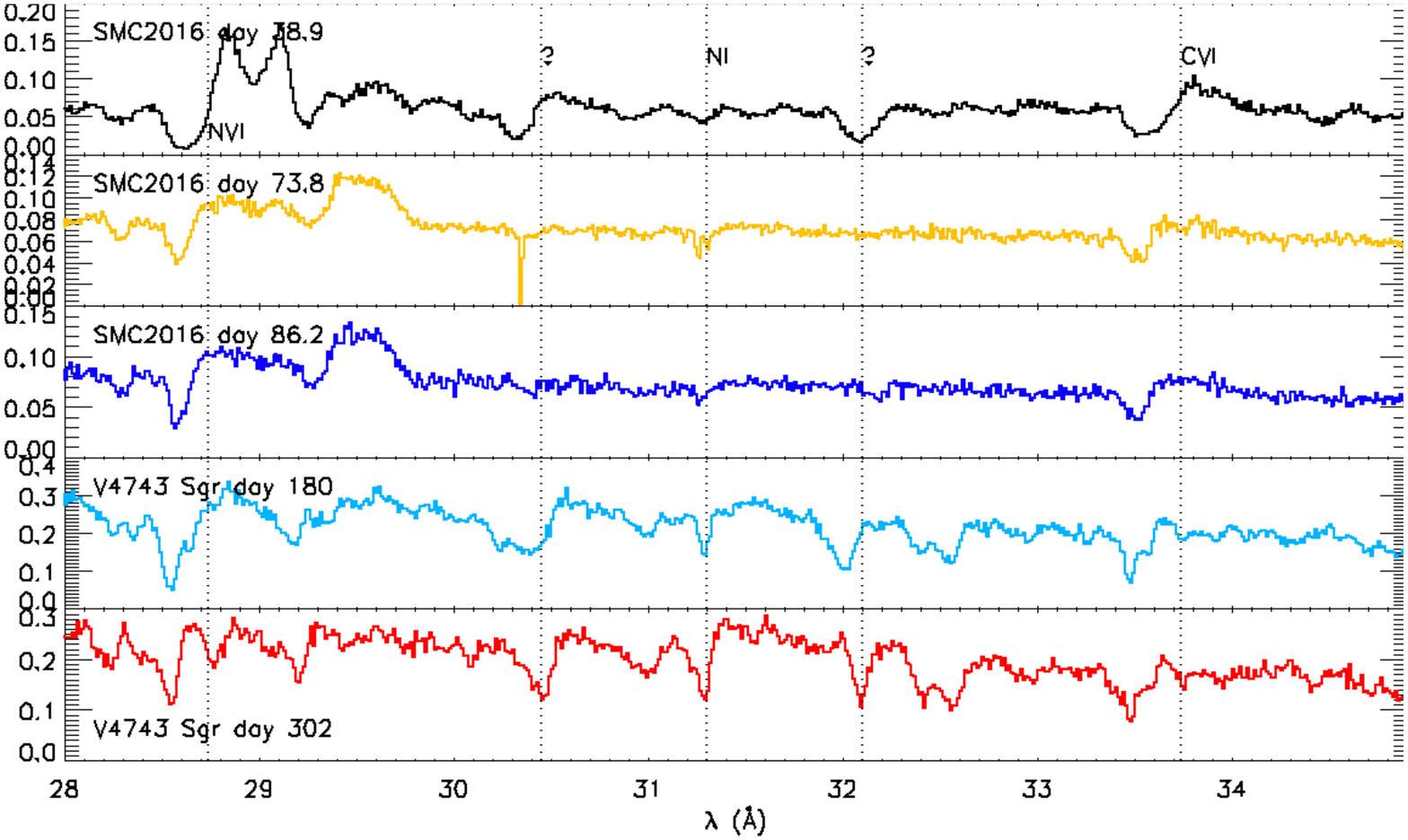}
        \caption{Evolution of absorption lines in the two
        novae SMC\,2016 (top three panels, {\it Chandra} ObsIDs
        19011/19012 (days 38.9 and 86.2)  and {\it XMM-Newton} ObsID 0794180201
        (day 73.8)) and
        V4743\,Sgr (bottom two panels, {\it Chandra} ObsIDs
        3775 (day 180) and 3776 (day 302)).
        Dotted lines mark a N\,{\sc vi} resonance line
        (28.8\,\AA), an unidentified line at $\sim 30.5$\,\AA,
        an interstellar N\,{\sc i} 1s-2p line at 31.3\,\AA, another
        unidentified line at $\sim 32.1$\,\AA, and a C\,{\sc vi}
        resonance line at 33.7\,\AA. The N\,{\sc vi} and C\,{\sc vi}
        lines are blue shifted by the same amount during the evolution
        indicating continuous expanding absorbing material
        (see \S~\ref{S:BlueShift}).
        The unidentified line at 32.1\,\AA\
        undergoes strange changes. In SMC\,2016, it was only
        present in the first observation while in V4743\,Sgr, it
        has changed from 32\,\AA\ to 32.1\,\AA\
        (see \S~\ref{S:UnidentLines}).}
        \label{unknown}
\end{figure}

\begin{figure}
%        \centering
        \includegraphics[scale=.37]{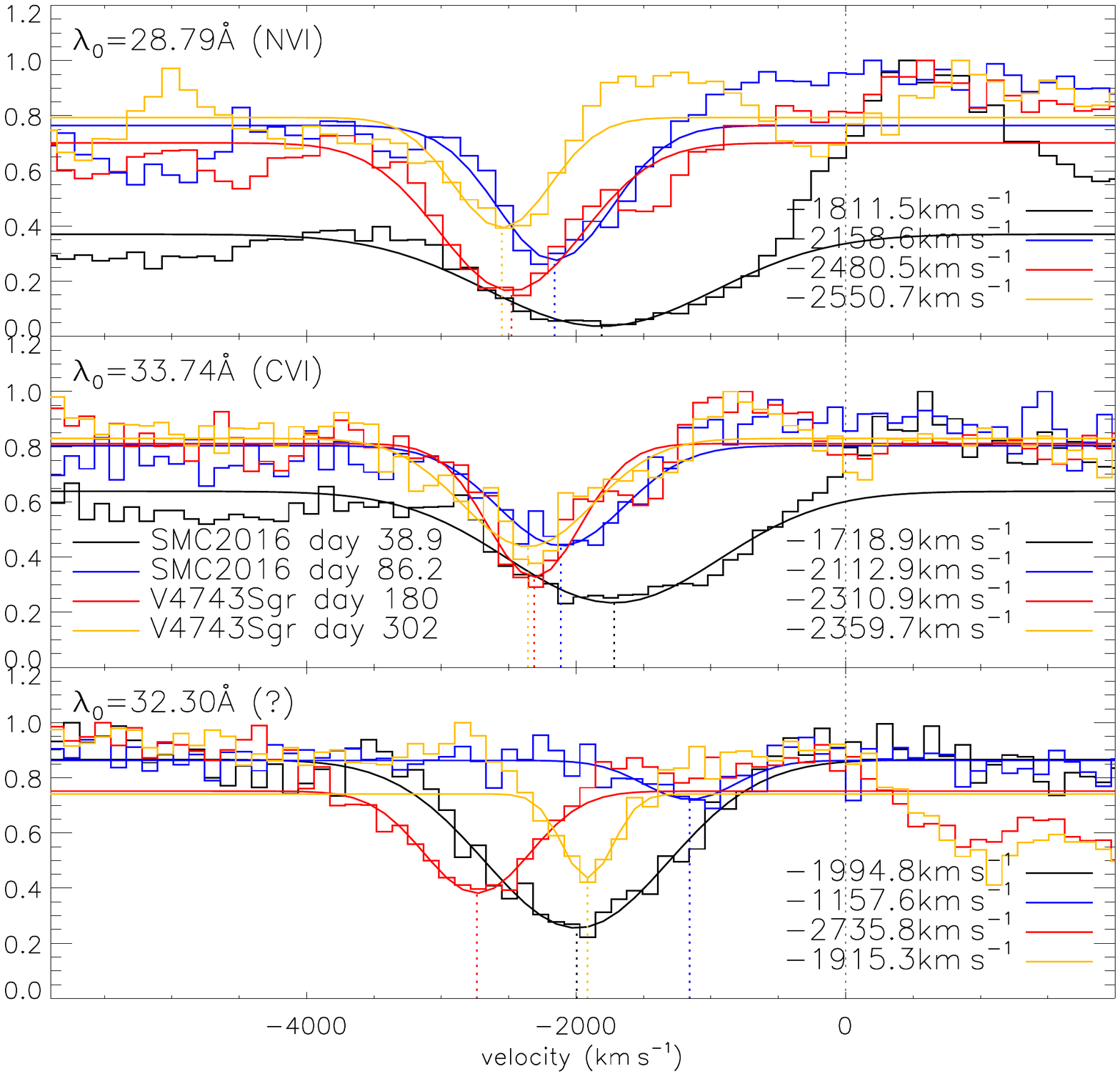}
        \caption{Comparison of line profiles of four spectra
	 shown in Fig.~\ref{unknown} (see text).}
        \label{lprof}
\end{figure}

 Already the first SSS spectrum of a nova, V4743\,Sgr, displayed absorption lines
 that are blue-shifted in excess of 2000\,km\,s$^{-1}$ \citep{v4743}.
 \cite{ness09} showed that line blue shifts are observed in many
 nova SSS spectra. In Fig.~\ref{unknown}, some example spectra are
 shown where the blue shifts can be seen in two resonance lines of
 N\,{\sc vi} and C\,{\sc vi}. We have determined the blue shifts
 with Gauss fits to the respective absorption lines and show the
 results in Fig.~\ref{lprof}. In SMC\,2016, the expansion velocity
 has increased from day 38.9 to day 86.2 (47 days) by $\sim 20$\% in
 both lines while in V4743\,Sgr, it has only increased, from day 180 to
 day 203 (122 days), by $\sim 3$\%.
 The increase of observed blue shifts may indicate that during
 the evolution of the SSS phase, the photospheric radius has continued
 to shrink into a regime with higher velocity, and the outflow is
 thus not constant with radius.\\
 
 The observed blue-shifted lines may either be formed in an optically
 thin wind or in the ejecta. In SMC\,2016, one can see evidence for
 P\,Cyg profiles in the first observation, whereas the emission
 component is not present in the later two observations. A similar
 behaviour was seen in RS\,Oph \citep{nessrsoph}. This may indicate
 that the mass loss rate can decrease during the SSS phase.\\

 Regardless of whether the absorption lines are formed in an optically
 thin wind or in the ejecta, both atmosphere model approaches compute
 the absorption lines as part of the same system. The model by
 \cite{rauch10} is a hydrostatic model that cannot reproduce blue
 shifts self-consistently, and in order
 to fit the blue shifts in the observed spectra, a negative
 red-shift parameter was added. All the dynamics are thus not included,
 and it is unknown what effect (apart from blue-shifted absorption
 lines) the expansion can have on the observable spectrum. It is also
 not confirmed whether the photosphere is already on the surface of the
 white dwarf or somewhere within the outflow.\\
 The wind-type model by \cite{vanrossum2012} assumes a hydrostatic core
 with an optically thin wind on top of that and models line blue shifts
 selfconsistently. The absorption behaviour of the wind can lead to
 significantly different model spectra compared to pure hydrostatic
 models, and adjustment to observed
 spectra can thus lead to much different parameters.\\
 It may be instructive to compare the \cite{rauch10} model with the
 hydrostatic core of the \cite{vanrossum2012} which should lead to
 consistent results if the different implementations and side assumptions
 are correct.

\subsection{Unidentified lines}
\label{S:UnidentLines}

The example spectra in Fig.~\ref{unknown} also show two unidentified
lines at $\sim 30.5$\,\AA\ and $\sim 32.1$\,\AA\ in both novae. In
SMC\,2016, both lines were only present on day 38.9 while in V4743\,Sgr,
they are seen in both example spectra, albeit with different profiles.
Strangely, one line shifted from 32\,\AA\ on day 180 to 32.1\,\AA\
on day 302. A line shift of 0.1\,\AA\ at 32\,\AA\ corresponds to
almost 1000\,km\,s$^{-1}$. Only two spectra are shown as examples,
but I looked at all five observation of V4743\,Sgr
({\it Chandra} ObsIDs 3775 (day 180), 3776 (day 302), 4435 (day 371),
and 5292 (day 526) and {\it XMM-Newton}
ObsID 0127720501 (day 196)) and found that this change already took
place between day 180 and day 196 (see bottom panel of Fig.~\ref{lprof_v4743})
- thus within only two weeks! - and
remained at 32.1\,\AA\ after day 196. We know nothing about this line,
it is obviously not interstellar and thus unlikely to arise from a
neutral atom like the N\,{\sc i} line at 31.3\,\AA. If it arises in
the same blue-shifted system as the high-ionisation lines of
N\,{\sc vi} and C\,{\sc vi}, the wavelength change in V4743\,Sgr
suggests a rest wavelength of around 32.3\,\AA. Various databases give
inconsistent results, e.g., there are several lines in NIST around
32.3\,\AA, e.g., Mg\,{\sc xi}, S\,{\sc xiii} while Atomdb\\
(http://www.atomdb.org/Webguide/webguide.php)\\
lists K\,{\sc xi} lines
that go to the ground. Such identifications need to be carefully checked
for consistency, e.g., with other lines in the same iso-electronic
sequence which is beyond of this work. There are numerous
other observed absorption lines without clear identifications in other
novae, see table~5 in \cite{nessv2491}. An approach to identify them
would be to assemble a data base with observed wavelengths of unknown
lines and the environments in which they are formed such as effective
temperature, blue shifts of known absorption lines etc. The identification
is important to improve the atomic data underlying the atmosphere models.

\subsection{Complex line profiles}

\begin{figure}
%        \centering
\includegraphics[scale=.37]{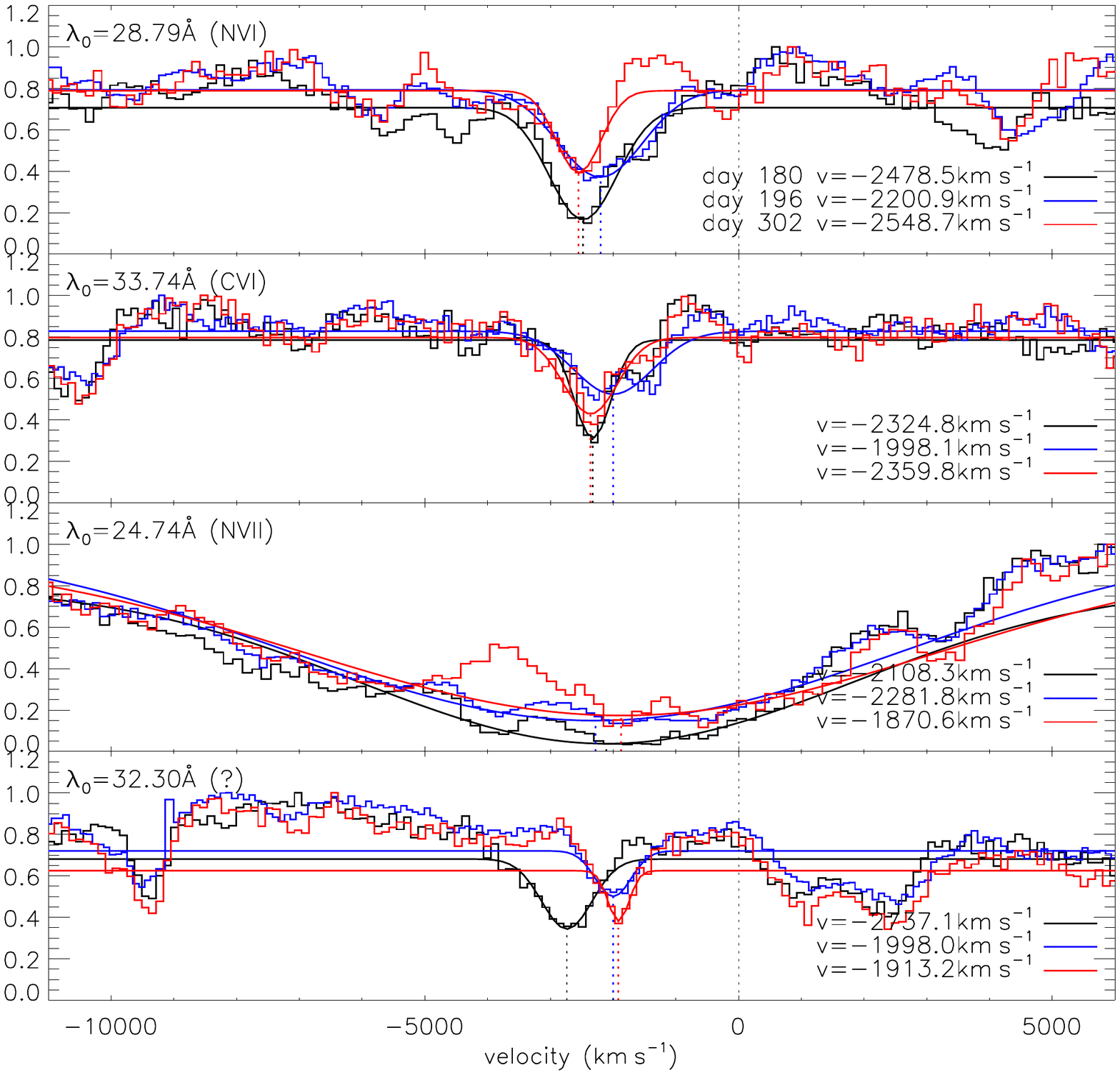}
        \caption{Comparison of line profiles of three spectra
	of nova V4743\,Sgr at different times of the SSS evolution
	(see labels).
	The top three panels compare the profiles of the three
	photospheric lines of N\,{\sc vi}, C\,{\sc vi}, and
	N\,{\sc vii}
	while the bottom panel shows the unidentified line
	assuming a rest wavelength of 32.3\,\AA.
	Best-fit blue shifts are given in the respective bottom
	right legends.
	 }
        \label{lprof_v4743}
\end{figure}

In a hydrostatic atmosphere, one expects narrow absorption lines,
but we see not only blue-shifted but also broadened absorption lines.
This is illustrated in Fig.~\ref{lprof_v4743} where the two photospheric
lines from Fig.~\ref{lprof} are shown in comparison to the N\,{\sc vii}
line at 24.74\,\AA. One can see that the profiles differ dramatically
with the N\,{\sc vii} line yielding a line width of several thousand
km\,s$^{-1}$ while the N\,{\sc vi} and C\,{\sc vi} lines are much
narrower. The line profile of the N\,{\sc vii} line is also
complex with at least two sub-components that have similar widths
as the N\,{\sc vi} and C\,{\sc vi} lines. One of these components,
at $\sim -4000$\,km\,s$^{-1}$,
has converted from an absorption line on day 180 to an emission
line on day 196 but then back to an absorption line day 302;
possibly also the lower-blueshift line at $\sim -2500$\,km\,s$^{-1}$.\\

The complexity of the N\,{\sc vii} line profile indicates that
we are observing several velocity components simultaneously.
This observation adds to numerous observational evidence for
non-homogeneous or asymmetric outflows from novae such as high-amplitude
variations during the early SSS phase \citep{osborne11}.
Therefore, even the model by \cite{vanrossum2012} is oversimplified.

\section{Conclusions}

The standard interpretation approach in X-ray astronomy for
spectra is to fit spectral models to the observations. When
the spectral resolution is low enough that no details are
resolved, this approach yields acceptable fits with already
quite simple models. If resulting parameters are unrealistic,
more complex models need to be tested, but it brings about
the problem that such models are overdetermined when fitted
to low-resolution spectra with only a few hundred spectral
bins.\\

With the advent of high-resolution X-ray spectra, a paradigm
shift would be necessary, but this has not been realised yet.\\

Highly complex atmosphere models provide of course more
physics than blackbody fits, however, they only give acceptable
fits when fitted to low-resolution spectra. No atmosphere model
has so far been found that reproduces an SSS grating spectra
in all the details. A quantitative assessment of the
goodness of fit appears hopeless at this time, and
the best that has been achieved was qualitative
agreement with data.\\

So, are the atmosphere models useless? And what can actually
be learned from the high-quality SSS grating spectra?\\

I propose to use the atmosphere models for the interpretation
of the data rather than trying to determine physical, quantitative
parameters. It is of course important to get a handle on the
effective temperature - but what does it help if we can never
be sure whether we have actually determed the correct value?
Rather than believing we have found the correct values, we
should consider any values found as an {\em assumption} because
they depend on the assumptions behind the models.\\

An example use case for the atmosphere models is the identification
of the absorption lines which is tricky because each identification
needs to be consistent with the rest of the spectrum. It is not
enough to consult an atomic database for the strongest line at a
given wavelength. Any candidate identification needs to be checked for
consistency, e.g., which other lines should then be seen or not seen.
Doing this manually is at best tedious because one would have
to go through the entire multiplet, finding the strongest lines
and compare which ones are seen at the expected wavelengths
and with the expected strengths. It would not be possible to
account for complex processes such as line emission and self
absorption (thus radiation transport), and
an atmosphere model would be the only possibility to determine
line identifications under complex conditions in a self-consistent
way.\\

When looking beyond X-ray astronomy, one will encounter
numerous methods of interpretations, e.g., of optical spectra, not
using any models at all. Optical spectra are traditionally
of too high quality that it was hopeless from the start to
find any model that could give quantitative parameter values.
Yet, a lot has been learned from optical spectra by classifying
spectra by presence or absence of certain lines, creating
classifications based on anomalously strong or weak types
of lines etc. The limited spectral resolution of X-ray
observations has not allowed such approaches so far, but
after almost 20 years of having high-resolution grating
spectra, it is time to explore more such methods. An
example is the classification of SSS spectra dominated
by emission or absorption lines by \cite{nessobsc}.\\

A way forward would be to study the diverse absorption line
profiles, identify the corresponding transitions and classify
lines behaving similarly in order to determine the dynamics in
the different regions of the outflow. The formation characteristics
of each line give physical information such as temperature, so
one can qualitatively localise the emission region, assuming a
radial temperature profile.\\

Future high-energy observatories such as XRISM or Athena
will provide more high-resolution spectra, and I predict
the days of model fitting to low-resolution spectra to be
counted.

\bibliographystyle{model2-names}

\bibliography{refs}

\begin{thebibliography}{24}
\expandafter\ifx\csname natexlab\endcsname\relax\def\natexlab#1{#1}\fi
\providecommand{\url}[1]{\texttt{#1}}
\providecommand{\href}[2]{#2}
\providecommand{\path}[1]{#1}
\providecommand{\DOIprefix}{doi:}
\providecommand{\ArXivprefix}{arXiv:}
\providecommand{\URLprefix}{URL: }
\providecommand{\Pubmedprefix}{pmid:}
\providecommand{\doi}[1]{\href{http://dx.doi.org/#1}{\path{#1}}}
\providecommand{\Pubmed}[1]{\href{pmid:#1}{\path{#1}}}
\providecommand{\bibinfo}[2]{#2}
\ifx\xfnm\relax \def\xfnm[#1]{\unskip,\space#1}\fi
%Type = Article
\bibitem[{{Balman} et~al.(1998){Balman}, {Krautter} and {\"Ogelman}}]{balm98}
\bibinfo{author}{{Balman}, S.}, \bibinfo{author}{{Krautter}, J.},
  \bibinfo{author}{{\"Ogelman}, H.}, \bibinfo{year}{1998}.
\newblock \bibinfo{title}{{The X-Ray Spectral Evolution of Classical Nova V1974
  Cygni 1992: A Reanalysis of the ROSAT Data}}.
\newblock \bibinfo{journal}{ApJ} \bibinfo{volume}{499},
  \bibinfo{pages}{395--406}.
%Type = Article
\bibitem[{{Gehrz} et~al.(2015){Gehrz}, {Evans}, {Helton}, {Shenoy}, {Banerjee},
  {Woodward}, {Vacca}, {Dykhoff}, {Ashok} and {Cass}}]{Gehrz2015}
\bibinfo{author}{{Gehrz}, R.D.}, \bibinfo{author}{{Evans}, A.},
  \bibinfo{author}{{Helton}, L.A.}, \bibinfo{author}{{Shenoy}, D.P.},
  \bibinfo{author}{{Banerjee}, D.P.K.}, \bibinfo{author}{{Woodward}, C.E.},
  \bibinfo{author}{{Vacca}, W.D.}, \bibinfo{author}{{Dykhoff}, D.A.},
  \bibinfo{author}{{Ashok}, N.M.}, \bibinfo{author}{{Cass}, A.C.},
  \bibinfo{year}{2015}.
\newblock \bibinfo{title}{{The Early Infrared Temporal Development of Nova
  Delphini 2013 (V339 DEL) Observed with the Stratospheric Observatory for
  Infrared Astronomy (SOFIA) and from the Ground}}.
\newblock \bibinfo{journal}{ApJ} \bibinfo{volume}{812},
  \bibinfo{pages}{132--143}.
\newblock \DOIprefix\doi{10.1088/0004-637X/812/2/132}.
%Type = Article
\bibitem[{{Gehrz} et~al.(1974){Gehrz}, {Hackwell} and {Jones}}]{ghj74}
\bibinfo{author}{{Gehrz}, R.D.}, \bibinfo{author}{{Hackwell}, J.A.},
  \bibinfo{author}{{Jones}, T.W.}, \bibinfo{year}{1974}.
\newblock \bibinfo{title}{{Infrared observations of Be stars from 2.3 to 19.5
  microns.}}
\newblock \bibinfo{journal}{ApJ} \bibinfo{volume}{191},
  \bibinfo{pages}{675--684}.
\newblock \DOIprefix\doi{10.1086/153008}.
%Type = Article
\bibitem[{{Henze} et~al.(2014){Henze}, {Ness}, {Darnley}, {Bode}, {Williams},
  {Shafter}, {Kato} and {Hachisu}}]{henze14}
\bibinfo{author}{{Henze}, M.}, \bibinfo{author}{{Ness}, J.U.},
  \bibinfo{author}{{Darnley}, M.J.}, \bibinfo{author}{{Bode}, M.F.},
  \bibinfo{author}{{Williams}, S.C.}, \bibinfo{author}{{Shafter}, A.W.},
  \bibinfo{author}{{Kato}, M.}, \bibinfo{author}{{Hachisu}, I.},
  \bibinfo{year}{2014}.
\newblock \bibinfo{title}{{A remarkable recurrent nova in M 31: The X-ray
  observations}}.
\newblock \bibinfo{journal}{A\&A} \bibinfo{volume}{563},
  \bibinfo{pages}{L8--L12}.
\newblock \DOIprefix\doi{10.1051/0004-6361/201423410},
  \href{http://arxiv.org/abs/1401.2904}{{\tt arXiv:1401.2904}}.
%Type = Article
\bibitem[{{Kahabka} and {van den Heuvel}(1997)}]{kahab}
\bibinfo{author}{{Kahabka}, P.}, \bibinfo{author}{{van den Heuvel}, E.P.J.},
  \bibinfo{year}{1997}.
\newblock \bibinfo{title}{{Luminous Supersoft X-Ray Sources}}.
\newblock \bibinfo{journal}{ARA\&A} \bibinfo{volume}{35},
  \bibinfo{pages}{69--100}.
%Type = Article
\bibitem[{{Krautter} et~al.(1996){Krautter}, {\"Ogelman}, {Starrfield},
  {Wichmann} and {Pfeffermann}}]{krautt96}
\bibinfo{author}{{Krautter}, J.}, \bibinfo{author}{{\"Ogelman}, H.},
  \bibinfo{author}{{Starrfield}, S.}, \bibinfo{author}{{Wichmann}, R.},
  \bibinfo{author}{{Pfeffermann}, E.}, \bibinfo{year}{1996}.
\newblock \bibinfo{title}{{ROSAT X-Ray Observations of Nova V1974 Cygni: The
  Rise and Fall of the Brightest Supersoft X-Ray Source}}.
\newblock \bibinfo{journal}{ApJ} \bibinfo{volume}{456},
  \bibinfo{pages}{788--797}.
%Type = Article
\bibitem[{{Lanz} et~al.(2005){Lanz}, {Telis}, {Audard}, {Paerels}, {Rasmussen}
  and {Hubeny}}]{lanz04}
\bibinfo{author}{{Lanz}, T.}, \bibinfo{author}{{Telis}, G.A.},
  \bibinfo{author}{{Audard}, M.}, \bibinfo{author}{{Paerels}, F.},
  \bibinfo{author}{{Rasmussen}, A.P.}, \bibinfo{author}{{Hubeny}, I.},
  \bibinfo{year}{2005}.
\newblock \bibinfo{title}{{Non-LTE Model Atmosphere Analysis of the Large
  Magellanic Cloud Supersoft X-Ray Source CAL 83}}.
\newblock \bibinfo{journal}{ApJ} \bibinfo{volume}{619},
  \bibinfo{pages}{517--526}.
%Type = Article
\bibitem[{{Lyke} et~al.(2002){Lyke}, {Kelley}, {Gehrz} and {Woodward}}]{lyke}
\bibinfo{author}{{Lyke}, J.E.}, \bibinfo{author}{{Kelley}, M.S.},
  \bibinfo{author}{{Gehrz}, R.D.}, \bibinfo{author}{{Woodward}, C.E.},
  \bibinfo{year}{2002}.
\newblock \bibinfo{title}{{Free-Free Turnover in Nova V4743 Sgr 2002 \#3}}.
\newblock \bibinfo{journal}{Bulletin of the American Astronomical Society}
  \bibinfo{volume}{34}, \bibinfo{pages}{1161}.
%Type = Article
\bibitem[{{Ness}(2010)}]{ness09}
\bibinfo{author}{{Ness}, J.}, \bibinfo{year}{2010}.
\newblock \bibinfo{title}{{Observational evidence for expansion in the SSS
  spectra of novae}}.
\newblock \bibinfo{journal}{Astronomische Nachrichten} \bibinfo{volume}{331},
  \bibinfo{pages}{179--182}.
\newblock \DOIprefix\doi{10.1002/asna.200911322},
  \href{http://arxiv.org/abs/0908.4549}{{\tt arXiv:0908.4549}}.
%Type = Article
\bibitem[{{Ness} et~al.(2012){Ness}, {Schaefer}, {Dobrotka}, {Sadowski},
  {Drake}, {Barnard}, {Talavera}, {Gonzalez-Riestra}, {Page}, {Hernanz}, {Sala}
  and {Starrfield}}]{nessusco}
\bibinfo{author}{{Ness}, J.}, \bibinfo{author}{{Schaefer}, B.E.},
  \bibinfo{author}{{Dobrotka}, A.}, \bibinfo{author}{{Sadowski}, A.},
  \bibinfo{author}{{Drake}, J.J.}, \bibinfo{author}{{Barnard}, R.},
  \bibinfo{author}{{Talavera}, A.}, \bibinfo{author}{{Gonzalez-Riestra}, R.},
  \bibinfo{author}{{Page}, K.L.}, \bibinfo{author}{{Hernanz}, M.},
  \bibinfo{author}{{Sala}, G.}, \bibinfo{author}{{Starrfield}, S.},
  \bibinfo{year}{2012}.
\newblock \bibinfo{title}{{From X-Ray Dips to Eclipse: Witnessing Disk
  Reformation in the Recurrent Nova U Sco}}.
\newblock \bibinfo{journal}{ApJ} \bibinfo{volume}{745},
  \bibinfo{pages}{43--58}.
\newblock \DOIprefix\doi{10.1088/0004-637X/745/1/43},
  \href{http://arxiv.org/abs/1105.2717}{{\tt arXiv:1105.2717}}.
%Type = Article
\bibitem[{{Ness} et~al.(2007){Ness}, {Starrfield}, {Beardmore}, {Bode},
  {Drake}, {Evans}, {Gehrz}, {Goad}, {Gonzalez-Riestra}, {Hauschildt},
  {Krautter}, {O'Brien}, {Osborne}, {Page}, {Sch\"onrich} and
  {Woodward}}]{nessrsoph}
\bibinfo{author}{{Ness}, J.}, \bibinfo{author}{{Starrfield}, S.},
  \bibinfo{author}{{Beardmore}, A.}, \bibinfo{author}{{Bode}, M.F.},
  \bibinfo{author}{{Drake}, J.J.}, \bibinfo{author}{{Evans}, A.},
  \bibinfo{author}{{Gehrz}, R.}, \bibinfo{author}{{Goad}, M.},
  \bibinfo{author}{{Gonzalez-Riestra}, R.}, \bibinfo{author}{{Hauschildt}, P.},
  \bibinfo{author}{{Krautter}, J.}, \bibinfo{author}{{O'Brien}, T.J.},
  \bibinfo{author}{{Osborne}, J.P.}, \bibinfo{author}{{Page}, K.L.},
  \bibinfo{author}{{Sch\"onrich}, R.}, \bibinfo{author}{{Woodward}, C.},
  \bibinfo{year}{2007}.
\newblock \bibinfo{title}{{The SSS phase of RS\,Ophiuchi observed with Chandra
  and XMM I. Data and preliminary Models}}.
\newblock \bibinfo{journal}{ApJ} \bibinfo{volume}{665},
  \bibinfo{pages}{1334--1348}.
\newblock \href{http://arxiv.org/abs/astro-ph/0705.1206}{{\tt
  arXiv:astro-ph/0705.1206}}.
%Type = Article
\bibitem[{{Ness} et~al.(2003){Ness}, {Starrfield}, {Burwitz}, {Wichmann},
  {Hauschildt}, {Drake}, {Wagner}, {Bond}, {Krautter}, {Orio}, {Hernanz},
  {Gehrz}, {Woodward}, {Butt}, {Mukai}, {Balman} and {Truran}}]{v4743}
\bibinfo{author}{{Ness}, J.}, \bibinfo{author}{{Starrfield}, S.},
  \bibinfo{author}{{Burwitz}, V.}, \bibinfo{author}{{Wichmann}, R.},
  \bibinfo{author}{{Hauschildt}, P.}, \bibinfo{author}{{Drake}, J.J.},
  \bibinfo{author}{{Wagner}, R.M.}, \bibinfo{author}{{Bond}, H.E.},
  \bibinfo{author}{{Krautter}, J.}, \bibinfo{author}{{Orio}, M.},
  \bibinfo{author}{{Hernanz}, M.}, \bibinfo{author}{{Gehrz}, R.D.},
  \bibinfo{author}{{Woodward}, C.E.}, \bibinfo{author}{{Butt}, Y.},
  \bibinfo{author}{{Mukai}, K.}, \bibinfo{author}{{Balman}, S.},
  \bibinfo{author}{{Truran}, J.W.}, \bibinfo{year}{2003}.
\newblock \bibinfo{title}{{A Chandra Low Energy Transmission Grating
  Spectrometer Observation of V4743 Sagittarii: A Supersoft X-Ray Source and a
  Violently Variable Light Curve}}.
\newblock \bibinfo{journal}{ApJL} \bibinfo{volume}{594},
  \bibinfo{pages}{L127--L130}.
%Type = Article
\bibitem[{{Ness} et~al.(2011){Ness}, {Osborne}, {Dobrotka}, {Page}, {Drake},
  {Pinto}, {Detmers}, {Schwarz}, {Bode}, {Beardmore}, {Starrfield}, {Hernanz},
  {Sala}, {Krautter} and {Woodward}}]{nessv2491}
\bibinfo{author}{{Ness}, J.U.}, \bibinfo{author}{{Osborne}, J.P.},
  \bibinfo{author}{{Dobrotka}, A.}, \bibinfo{author}{{Page}, K.L.},
  \bibinfo{author}{{Drake}, J.J.}, \bibinfo{author}{{Pinto}, C.},
  \bibinfo{author}{{Detmers}, R.G.}, \bibinfo{author}{{Schwarz}, G.},
  \bibinfo{author}{{Bode}, M.F.}, \bibinfo{author}{{Beardmore}, A.P.},
  \bibinfo{author}{{Starrfield}, S.}, \bibinfo{author}{{Hernanz}, M.},
  \bibinfo{author}{{Sala}, G.}, \bibinfo{author}{{Krautter}, J.},
  \bibinfo{author}{{Woodward}, C.E.}, \bibinfo{year}{2011}.
\newblock \bibinfo{title}{{XMM-Newton X-ray and Ultraviolet Observations of the
  Fast Nova V2491 Cyg during the Supersoft Source Phase}}.
\newblock \bibinfo{journal}{ApJ} \bibinfo{volume}{733},
  \bibinfo{pages}{70--85}.
\newblock \DOIprefix\doi{10.1088/0004-637X/733/1/70},
  \href{http://arxiv.org/abs/1103.4543}{{\tt arXiv:1103.4543}}.
%Type = Article
\bibitem[{{Ness} et~al.(2013){Ness}, {Osborne}, {Henze}, {Dobrotka}, {Drake},
  {Ribeiro}, {Starrfield}, {Kuulkers}, {Behar}, {Hernanz}, {Schwarz}, {Page},
  {Beardmore} and {Bode}}]{nessobsc}
\bibinfo{author}{{Ness}, J.U.}, \bibinfo{author}{{Osborne}, J.P.},
  \bibinfo{author}{{Henze}, M.}, \bibinfo{author}{{Dobrotka}, A.},
  \bibinfo{author}{{Drake}, J.J.}, \bibinfo{author}{{Ribeiro}, V.A.R.M.},
  \bibinfo{author}{{Starrfield}, S.}, \bibinfo{author}{{Kuulkers}, E.},
  \bibinfo{author}{{Behar}, E.}, \bibinfo{author}{{Hernanz}, M.},
  \bibinfo{author}{{Schwarz}, G.}, \bibinfo{author}{{Page}, K.L.},
  \bibinfo{author}{{Beardmore}, A.P.}, \bibinfo{author}{{Bode}, M.F.},
  \bibinfo{year}{2013}.
\newblock \bibinfo{title}{{Obscuration effects in super-soft-source X-ray
  spectra}}.
\newblock \bibinfo{journal}{A\&A} \bibinfo{volume}{559},
  \bibinfo{pages}{1--15}.
\newblock \DOIprefix\doi{10.1051/0004-6361/201322415},
  \href{http://arxiv.org/abs/1309.2604}{{\tt arXiv:1309.2604}}.
%Type = Article
\bibitem[{{Osborne} et~al.(2011){Osborne}, {Page}, {Beardmore}, {Bode}, {Goad},
  {O'Brien}, {Starrfield}, {Rauch}, {Ness}, {Krautter}, {Schwarz}, {Burrows},
  {Gehrels}, {Drake}, {Evans} and {Eyres}}]{osborne11}
\bibinfo{author}{{Osborne}, J.P.}, \bibinfo{author}{{Page}, K.L.},
  \bibinfo{author}{{Beardmore}, A.P.}, \bibinfo{author}{{Bode}, M.F.},
  \bibinfo{author}{{Goad}, M.R.}, \bibinfo{author}{{O'Brien}, T.J.},
  \bibinfo{author}{{Starrfield}, S.}, \bibinfo{author}{{Rauch}, T.},
  \bibinfo{author}{{Ness}, J.}, \bibinfo{author}{{Krautter}, J.},
  \bibinfo{author}{{Schwarz}, G.}, \bibinfo{author}{{Burrows}, D.N.},
  \bibinfo{author}{{Gehrels}, N.}, \bibinfo{author}{{Drake}, J.J.},
  \bibinfo{author}{{Evans}, A.}, \bibinfo{author}{{Eyres}, S.P.S.},
  \bibinfo{year}{2011}.
\newblock \bibinfo{title}{{The Supersoft X-ray Phase of Nova RS Ophiuchi
  2006}}.
\newblock \bibinfo{journal}{ApJ} \bibinfo{volume}{727},
  \bibinfo{pages}{124--123}.
\newblock \DOIprefix\doi{10.1088/0004-637X/727/2/124},
  \href{http://arxiv.org/abs/1011.5327}{{\tt arXiv:1011.5327}}.
%Type = Article
\bibitem[{{Parmar} et~al.(1997a){Parmar}, {Kahabka}, {Hartmann}, {Heise},
  {Martin}, {Bavdaz} and {Mineo}}]{parmarcal87}
\bibinfo{author}{{Parmar}, A.N.}, \bibinfo{author}{{Kahabka}, P.},
  \bibinfo{author}{{Hartmann}, H.W.}, \bibinfo{author}{{Heise}, J.},
  \bibinfo{author}{{Martin}, D.D.E.}, \bibinfo{author}{{Bavdaz}, M.},
  \bibinfo{author}{{Mineo}, T.}, \bibinfo{year}{1997}a.
\newblock \bibinfo{title}{{A BeppoSAX observation of the super-soft source
  CAL87.}}
\newblock \bibinfo{journal}{A\&A} \bibinfo{volume}{323},
  \bibinfo{pages}{L33--L36}.
\newblock \href{http://arxiv.org/abs/arXiv:astro-ph/9706008}{{\tt
  arXiv:arXiv:astro-ph/9706008}}.
%Type = Article
\bibitem[{{Parmar} et~al.(1998){Parmar}, {Kahabka}, {Hartmann}, {Heise} and
  {Taylor}}]{parmarcal83}
\bibinfo{author}{{Parmar}, A.N.}, \bibinfo{author}{{Kahabka}, P.},
  \bibinfo{author}{{Hartmann}, H.W.}, \bibinfo{author}{{Heise}, J.},
  \bibinfo{author}{{Taylor}, B.G.}, \bibinfo{year}{1998}.
\newblock \bibinfo{title}{{A BeppoSAX LECS observation of the super-soft X-ray
  source CAL 83}}.
\newblock \bibinfo{journal}{A\&A} \bibinfo{volume}{332},
  \bibinfo{pages}{199--203}.
\newblock \href{http://arxiv.org/abs/arXiv:astro-ph/9712040}{{\tt
  arXiv:arXiv:astro-ph/9712040}}.
%Type = Article
\bibitem[{{Parmar} et~al.(1997b){Parmar}, {Martin}, {Bavdaz}, {Favata},
  {Kuulkers}, {Vacanti}, {Lammers}, {Peacock} and {Taylor}}]{parmar97}
\bibinfo{author}{{Parmar}, A.N.}, \bibinfo{author}{{Martin}, D.D.E.},
  \bibinfo{author}{{Bavdaz}, M.}, \bibinfo{author}{{Favata}, F.},
  \bibinfo{author}{{Kuulkers}, E.}, \bibinfo{author}{{Vacanti}, G.},
  \bibinfo{author}{{Lammers}, U.}, \bibinfo{author}{{Peacock}, A.},
  \bibinfo{author}{{Taylor}, B.G.}, \bibinfo{year}{1997}b.
\newblock \bibinfo{title}{{The low-energy concentrator spectrometer on-board
  the BeppoSAX X-ray astronomy satellite}}.
\newblock \bibinfo{journal}{AAPS} \bibinfo{volume}{122},
  \bibinfo{pages}{309--326}.
\newblock \DOIprefix\doi{10.1051/aas:1997137}.
%Type = Article
\bibitem[{{Rauch} et~al.(2010){Rauch}, {Orio}, {Gonzales-Riestra}, {Nelson},
  {Still}, {Werner} and {Wilms}}]{rauch10}
\bibinfo{author}{{Rauch}, T.}, \bibinfo{author}{{Orio}, M.},
  \bibinfo{author}{{Gonzales-Riestra}, R.}, \bibinfo{author}{{Nelson}, T.},
  \bibinfo{author}{{Still}, M.}, \bibinfo{author}{{Werner}, K.},
  \bibinfo{author}{{Wilms}, J.}, \bibinfo{year}{2010}.
\newblock \bibinfo{title}{{Non-local Thermal Equilibrium Model Atmospheres for
  the Hottest White Dwarfs: Spectral Analysis of the Compact Component in Nova
  V4743 Sgr}}.
\newblock \bibinfo{journal}{ApJ} \bibinfo{volume}{717},
  \bibinfo{pages}{363--371}.
\newblock \DOIprefix\doi{10.1088/0004-637X/717/1/363},
  \href{http://arxiv.org/abs/1006.2918}{{\tt arXiv:1006.2918}}.
%Type = Article
\bibitem[{{Rauch} et~al.(2008){Rauch}, {Suleimanov} and {Werner}}]{rauch08}
\bibinfo{author}{{Rauch}, T.}, \bibinfo{author}{{Suleimanov}, V.},
  \bibinfo{author}{{Werner}, K.}, \bibinfo{year}{2008}.
\newblock \bibinfo{title}{{Absorption features in the spectra of X-ray bursting
  neutron stars}}.
\newblock \bibinfo{journal}{A\&A} \bibinfo{volume}{490},
  \bibinfo{pages}{1127--1134}.
\newblock \DOIprefix\doi{10.1051/0004-6361:200810129},
  \href{http://arxiv.org/abs/0809.2170}{{\tt arXiv:0809.2170}}.
%Type = Article
\bibitem[{{Schwarz} et~al.(2011){Schwarz}, {Ness}, {Osborne}, {Page}, {Evans},
  {Beardmore}, {Walter}, {Helton}, {Woodward}, {Bode}, {Starrfield} and
  {Drake}}]{schwarz2011}
\bibinfo{author}{{Schwarz}, G.J.}, \bibinfo{author}{{Ness}, J.},
  \bibinfo{author}{{Osborne}, J.P.}, \bibinfo{author}{{Page}, K.L.},
  \bibinfo{author}{{Evans}, P.A.}, \bibinfo{author}{{Beardmore}, A.P.},
  \bibinfo{author}{{Walter}, F.M.}, \bibinfo{author}{{Helton}, L.A.},
  \bibinfo{author}{{Woodward}, C.E.}, \bibinfo{author}{{Bode}, M.},
  \bibinfo{author}{{Starrfield}, S.}, \bibinfo{author}{{Drake}, J.J.},
  \bibinfo{year}{2011}.
\newblock \bibinfo{title}{{Swift X-Ray Observations of Classical Novae. II. The
  Super Soft Source Sample}}.
\newblock \bibinfo{journal}{ApJS} \bibinfo{volume}{197},
  \bibinfo{pages}{31--56}.
\newblock \DOIprefix\doi{10.1088/0067-0049/197/2/31},
  \href{http://arxiv.org/abs/1110.6224}{{\tt arXiv:1110.6224}}.
%Type = Article
\bibitem[{{van den Heuvel} et~al.(1992){van den Heuvel}, {Bhattacharya},
  {Nomoto} and {Rappaport}}]{heuvel}
\bibinfo{author}{{van den Heuvel}, E.P.J.}, \bibinfo{author}{{Bhattacharya},
  D.}, \bibinfo{author}{{Nomoto}, K.}, \bibinfo{author}{{Rappaport}, S.A.},
  \bibinfo{year}{1992}.
\newblock \bibinfo{title}{{Accreting white dwarf models for CAL 83, CAL 87 and
  other ultrasoft X-ray sources in the LMC}}.
\newblock \bibinfo{journal}{A\&A} \bibinfo{volume}{262},
  \bibinfo{pages}{97--105}.
%Type = Article
\bibitem[{{van Rossum}(2012)}]{vanrossum2012}
\bibinfo{author}{{van Rossum}, D.R.}, \bibinfo{year}{2012}.
\newblock \bibinfo{title}{{A Public Set of Synthetic Spectra from Expanding
  Atmospheres for X-Ray Novae. I. Solar Abundances}}.
\newblock \bibinfo{journal}{ApJ} \bibinfo{volume}{756},
  \bibinfo{pages}{43--53}.
\newblock \DOIprefix\doi{10.1088/0004-637X/756/1/43},
  \href{http://arxiv.org/abs/1205.4267}{{\tt arXiv:1205.4267}}.
%Type = Article
\bibitem[{{Werner} et~al.(2012){Werner}, {Rauch}, {Ringat} and
  {Kruk}}]{werner12}
\bibinfo{author}{{Werner}, K.}, \bibinfo{author}{{Rauch}, T.},
  \bibinfo{author}{{Ringat}, E.}, \bibinfo{author}{{Kruk}, J.W.},
  \bibinfo{year}{2012}.
\newblock \bibinfo{title}{{First Detection of Krypton and Xenon in a White
  Dwarf}}.
\newblock \bibinfo{journal}{ApJL} \bibinfo{volume}{753},
  \bibinfo{pages}{L7--L11}.
\newblock \DOIprefix\doi{10.1088/2041-8205/753/1/L7}.

\end{thebibliography}

\end{document}